\documentclass{llncs}
\usepackage{epsfig}
\title{{\bf Microwave absorption by the Josephson-junction network 
       in a low field: 
       A realistic model for ceramic high-temperature superconductors}}
\author{Adam Rycerz and Jozef Spa{\l}ek}
\institute{Marian Smoluchowski Institute of Physics,\\
        Jagiellonian University, ulica Reymonta~4,\\ 
        30-059 Krak\'ow, Poland}
\date{}

\def\req#1{(\ref{#1})}
\def\refig#1{Fig.\ref{#1}}
\newcommand{\duzy}[1]{{\textstyle #1}}
\newcommand{\duzyfrac}[2]{{\frac{\duzy{#1}}{\duzy{#2}}}}

\begin{document}
\pagestyle{plain}
\maketitle

\begin{abstract}
We discuss 
the applied magnetic field dependence of the absorption of microwaves by
a 3-dimensional array up to 30x30x30 Josephson junctions
with random parameters including the resistivity, capacity and inductance
of each junction. The numerical simulation results for the networks show
characteristic microwave absorption anomalies observed in the ceramic 
samples of high temperature
superconductor $\mbox{YBa}_2\mbox{Cu}_3\mbox{O}_{7-x}$. 
We also provide a discussion of the absorption in simple 
analytical terms of Josephson loop instabilities. \\

\end{abstract}

\section{Introduction}

\begin{figure*}
\centering\epsfxsize=0.8\textwidth\epsfbox{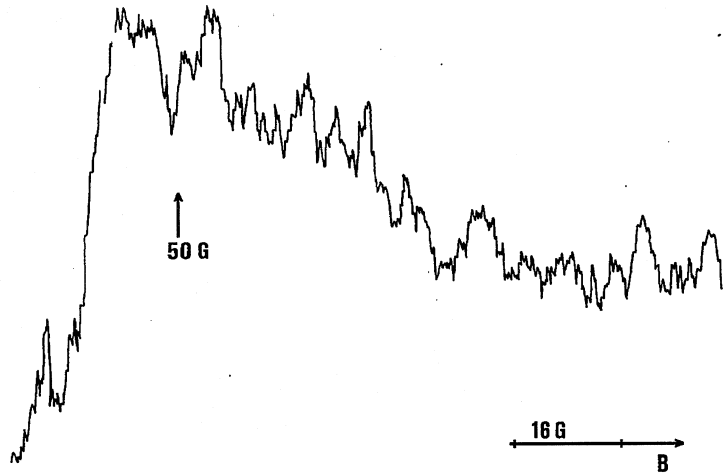}
\caption{
  First derivative the EPR power absorption showing the 50-G peak for sample 
  of $\mbox{YBa}_2\mbox{Cu}_3\mbox{O}_{6+x}$ at $T=80\mbox{ K}$.
  Reprinted from Ref. \cite{stan}.}
\label{abs-stan}
\end{figure*}

Soon after the discovery of high 
temperature superconductivity it has been established 
[1-3] that in ceramic samples of
$\mbox{YBa}_2\mbox{Cu}_3\mbox{O}_{7-x}$ 
there exists a~large nonresonant microwave 
absorption in zero and low applied magnetic field 
that can be associated with the transition to the 
superconducting phase. An exemplary field 
dependence of this absorption is displayed in 
\refig{abs-stan} \cite{stan}. A broad absorption structure is modulated by
a noisy signal with a clear evidence of 
a quasiperiodic substructure. One can 
also observe nonlinear phenomena, e.g. a generation 
of odd harmonics in zero applied field and  
of even harmonics in nonzero field \cite{jeff}. 

The experimental setup used to measure the spectrum 
depicted in \refig{abs-stan} is an EPR spectrometer, 
which provides the magnetic resonance response 
due to $\mbox{Cu}^{2+}$ ions in the normal phase in 
the field range $0.35\sim 0.4\ T$. The lower
field signal has a reversed phase with respect 
to the usual (Lorentzian) EPR line. Therefore, 
this low-field absorption has been attributed 
to the power loss in current loops connected 
by the dissipative Josephson 
junctions, which in turn are
created between the grains in a ceramic 
sample. Actually, such junctions appear also in the
monocrystaline sample \cite{vich}, although their presence
in good quality untwinned monocrystalline samples should
be associated with the Cooper-pair tunneling along the $c$-axis.

Because of the potential application of ceramic high temperature
superconductors in various microwave applications, 
it is important to study the properties 
of a general three-dimensional array of  Josephson junctions
shunted by resistance ($R$), capacitance ($C$), 
and with inductance ($L$), together with 
inclusion of some random variations of the junction parameters. 
Such model can be called the disordered {\em RLC-SJJ} model \cite{intro}.
For that purpose we consider a lattice depicted in
\refig{siec-th}, where ${\theta}^{\alpha}_{ijk}$ labels the phase difference
in the direction $\alpha=x,y,z$, and $(i,j,k)$ locate 
the junction position in the three-dimensional array. This model 
was considered in various simplified situations 
by a number of authors \cite{clem}. 
Here we consider a 3-dimensional granular structure and
disregard the circumstance that the grains 
themselves may have a coupled Josephson structure 
along the $c$-axis, perpendicular to the $\mbox{CuO}_2$ planes \cite{law}.
In other words, the intergranular 
critical currents are assumed to be much smaller 
than their intragranular counterpart. This should 
allow us to address the question about role of the Josephson network 
in causing the hysteresis when cycling slowly the 
applied field, as well as the question of the
existence of a residual contribution in the zero field.

\section{The model}
\subsection{Dissipative lattice of Josephson junctions}

\begin{figure*}[t]
\centering\epsfxsize=0.8\textwidth\epsfbox{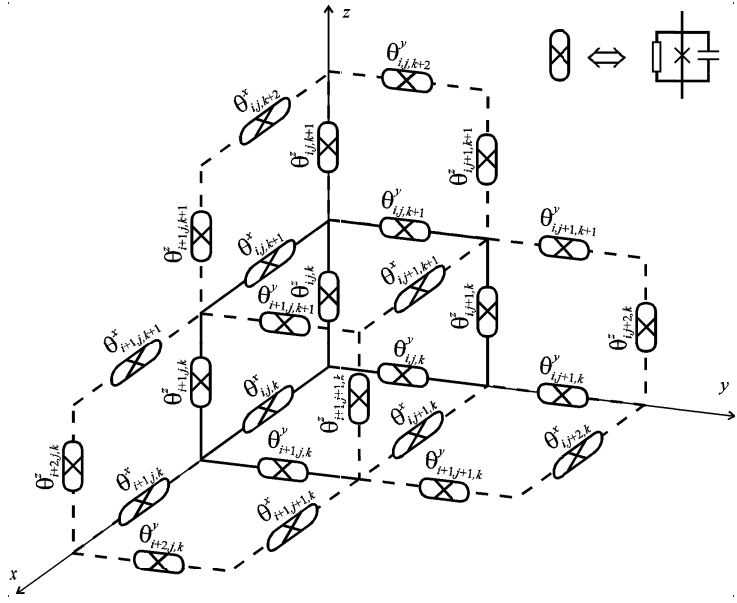}
\caption{
  The network with gauge invariant phases marked on each 
  Josephson junction. The $\theta^{\alpha}$ values grow 
  in the positive directions of the coresponding axis $\alpha=x,y,z$.}
\label{siec-th}
\end{figure*}

\begin{figure*}
\centering\epsfxsize=0.8\textwidth\epsfbox{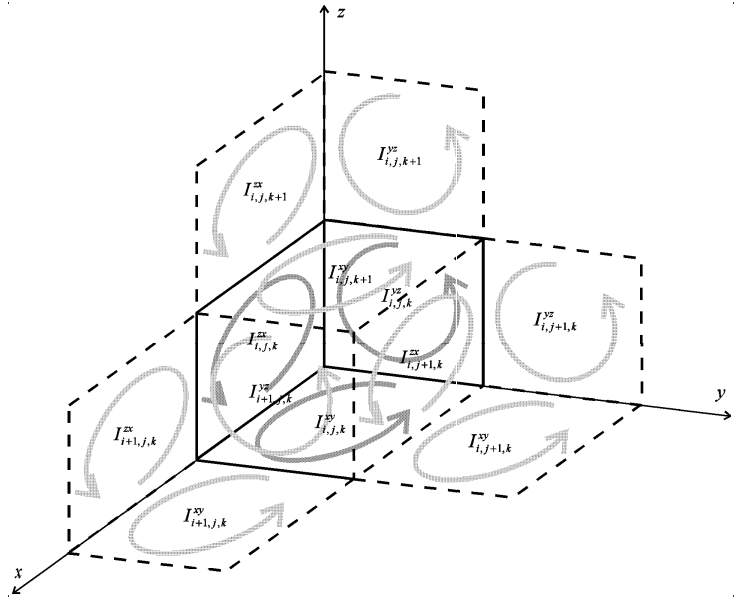}
\caption{
  The current distribution in loops composing 
  the lattice of the Josephson junctions.}
\label{siec-i}
\end{figure*}

We start from the gauge invariant 
phase $\theta$ between points $1$ and $2$: 
  \begin{equation}
  \label{g-inv}
    \theta_{12}\equiv\int_{1}^{2}d\vec{l}\cdot
    \left( \nabla\phi-\frac{2\pi}{\Phi_0}\vec{A} \right),
  \end{equation}
where $\nabla\phi$ is the gradient of the phase 
of macroscopic wave function, $\vec{A}$ in the 
vector potential, and the integration takes 
place between the two sides of the junction 
with the sign convention in accordance 
with the coordinate directions shown in \refig{siec-th}. Taking
into account the single valuedness of the wave 
function, the Stokes theorem for the elementary loop, we 
obtain the relation between e.~g. the magnetic flux 
piercing the loop in $xy$ plane and the 
phase differences on the Josephson junctions composing 
the loop in the form: 
  \begin{equation}
  \label{phi-xy}
    \frac{2\pi}{\Phi_0}\Phi^{xy}_{i,j,k}=
    -\left(
      \theta^{x}_{i,j,k}-\theta^{x}_{i,j+1,k}
      +\theta^{y}_{i+1,j,k}-\theta^{y}_{i,j,k}
    \right).
  \end{equation}
For the sake of convenience we label 
the loops currents $I^{\alpha\beta}_{ijk}$ in the same manner 
as the fluxes $\Phi^{\alpha\beta}_{ijk}$, as shown explicitly 
in \refig{siec-i}. 

Introducing reduced quantities 
$\tilde{\Phi}=\Phi/\Phi_0$, and $\tilde{\theta}=\theta/2\pi$ we can write 
down the relation \req{phi-xy} for each plane 
$\alpha\beta=xy,yz\mbox{ and }zx$ with the 
replacement $\Phi\rightarrow\tilde{\Phi}$ 
and $\theta\rightarrow\tilde{\theta}$, so that: 
  \begin{equation}
  \label{phi-ijk}
   \left\{\begin{array}{c}
      \tilde{\Phi}^{xy}_{i,j,k}=
      -\left(
        \tilde{\theta}^{x}_{i,j,k}+\tilde{\theta}^{y}_{i+1,j,k}
        -\tilde{\theta}^{x}_{i,j+1,k}-\tilde{\theta}^{y}_{i,j,k}
      \right)\\
      \tilde{\Phi}^{zx}_{i,j,k}=
      -\left(
        \tilde{\theta}^{z}_{i,j,k}+\tilde{\theta}^{x}_{i,j,k+1}
           -\tilde{\theta}^{z}_{i+1,j,k}-\tilde{\theta}^{x}_{i,j,k} 
      \right) \\
      \tilde{\Phi}^{yz}_{i,j,k}=
      -\left(
        \tilde{\theta}^{y}_{i,j,k}+\tilde{\theta}^{z}_{i,j+1,k}
        -\tilde{\theta}^{y}_{i,j,k+1}-\tilde{\theta}^{z}_{i,j,k} 
      \right)
    \end{array}\right. .
  \end{equation}
The cyclic properties of these relations are 
easy to grasp. The $(i,j,k)$ take 
the following values for the cubic 
array containing $N\times N\times N$ junctions: 
  \begin{equation}
  \label{ijk-fi}
    \left\{\begin{array}{llll}
      \tilde{\Phi}^{xy}_{i,j,k}: &  i=0,..,N-1; & j=0,..,N-1; & k=0,..,N;\\
      \tilde{\Phi}^{zx}_{i,j,k}: &  i=0,..,N-1; & j=0,..,N; & k=0,..,N-1;\\
      \tilde{\Phi}^{yz}_{i,j,k}: &  i=0,..,N; & j=0,..,N-1; & k=0,..,N-1.  
    \end{array}\right. 
  \end{equation}
The corresponding relation for the junction phase differences are:
  \begin{equation}
  \label{ijk-th} 
    \left\{\begin{array}{llll}
      \tilde{\theta}^{x}_{i,j,k}: & i=0,..,N-1; & j=0,..,N; & k=0,..,N;\\
      \tilde{\theta}^{y}_{i,j,k}: & i=0,..,N; & j=0,..,N-1; & k=0,..,N;\\
      \tilde{\theta}^{z}_{i,j,k}: & i=0,..,N; & j=0,..,N; & k=0,..,N-1.
    \end{array}\right. 
  \end{equation}

Let us write down the dynamic equation for the current
in individual loop. For a single 
$RC$ shunted Josephson junction (cf. the inset in \refig{siec-th}) 
the total current $I$ through it is composed of 3 terms, namely
  \begin{equation} 
  \label{r-rcsj}
    I=C\frac{\Phi_0}{2\pi}\ddot{\theta}+
    \frac{1}{R}\frac{\Phi_0}{2\pi}\dot{\theta}+I_c\sin\theta. 
  \end{equation}
where $C$ and $R$ are the junction 
capacity and resistance, respectively and $I_c$ is its
critical current (the self-inductance $L$ is included through the 
relation to the flux, $\Phi=-LI$, see below). In the case
of a junction placed on $z$ axis inside the network, 
the current through it is composed of four 
contributions from the loops attached to it, namely
  \begin{equation}
  \label{thz0}
    C\frac{\Phi_0}{2\pi}\ddot{\theta}^{z}_{i,j,k}+
    \frac{1}{R}\frac{\Phi_0}{2\pi}\dot{\theta}^{z}_{i,j,k}+
    I_c\sin\theta^{z}_{i,j,k}=
    I^{zx}_{i,j,k}-I^{zx}_{i-1,j,k}-I^{yz}_{i,j,k}+I^{yz}_{i,j-1,k}. 
  \end{equation}
Defining additional dimensionless variables
  \begin{equation}
  \label{bezwym2}
    \left\{\begin{array}{ccc}
      \tilde{I}\equiv I/I_c, & 
        \tilde{L}\equiv LI_c/\Phi_0, & 
        \tilde{\Gamma}\equiv\sqrt{LC}/RC, \\
      \tilde{t}\equiv t/\sqrt{LC}, &
        \dot{\tilde{\theta}}\equiv d\tilde{\theta}/d\tilde{t}, &
        \ddot{\tilde{\theta}}\equiv
          d^2\tilde{\theta}/d{\tilde{t}}^2, 
    \end{array}\right.
  \end{equation}
we can write down a complete set 
of dynamic equations for $\tilde{\theta}^{x,y,z}$ in the form
  \begin{equation}
  \label{th-xyz}
    \left\{\begin{array}{c}
      \ddot{\tilde{\theta}}^{x}_{i,j,k}+
      \tilde{\Gamma}\dot{\tilde{\theta}}^{x}_{i,j,k}+
      \tilde{L}\sin 2\pi\tilde{\theta}^{x}_{i,j,k}=
        \tilde{L}\left(
        \tilde{I}^{xy}_{i,j,k}-\tilde{I}^{xy}_{i,j-1,k}-
        \tilde{I}^{zx}_{i,j,k}+\tilde{I}^{zx}_{i,j,k-1} 
        \right) \\
      \ddot{\tilde{\theta}}^{y}_{i,j,k}+
      \tilde{\Gamma}\dot{\tilde{\theta}}^{y}_{i,j,k}+
      \tilde{L}\sin 2\pi\tilde{\theta}^{y}_{i,j,k}=
        \tilde{L}\left(
        \tilde{I}^{yz}_{i,j,k}-\tilde{I}^{yz}_{i,j,k-1}-
        \tilde{I}^{xy}_{i,j,k}+\tilde{I}^{xy}_{i-1,j,k} 
        \right) \\
      \ddot{\tilde{\theta}}^{z}_{i,j,k}+
      \tilde{\Gamma}\dot{\tilde{\theta}}^{z}_{i,j,k}+
      \tilde{L}\sin 2\pi\tilde{\theta}^{z}_{i,j,k}=
        \tilde{L}\left(
        \tilde{I}^{zx}_{i,j,k}-\tilde{I}^{zx}_{i-1,j,k}-
        \tilde{I}^{yz}_{i,j,k}+\tilde{I}^{yz}_{i,j-1,k} 
        \right) \\
    \end{array}\right. .
  \end{equation}
The boundary conditions are set as
  \begin{equation}
  \label{warb}
    \tilde{L}
    \left(\begin{array}{c}
      \tilde{I}^{xy}_{i,j,k} \\
      \tilde{I}^{zx}_{i,j,k} \\
      \tilde{I}^{yz}_{i,j,k} 
    \end{array}\right)=
    \left\{\begin{array}{cl} 
      \left(\begin{array}{c}
        \tilde{\Phi}^{xy}_{i,j,k}-\tilde{\Phi}^{ext}_{z} \\
        \tilde{\Phi}^{zx}_{i,j,k}-\tilde{\Phi}^{ext}_{y} \\
        \tilde{\Phi}^{yz}_{i,j,k}-\tilde{\Phi}^{ext}_{x} 
      \end{array}\right) &
       \mbox{for } (i,j,k) \in\req{ijk-fi}\\
      0 & \mbox{otherwise,}
    \end{array}\right. 
  \end{equation}
where $\tilde{\Phi}^{ext}_{\alpha}$ is the external flux 
(in units $\Phi_0$) enclosed by the plane 
perpendicular to the axis $\alpha=x,y,z$. Thus we see 
that the external flux is screened by the self-inductance 
and Josephson supercurrents at the surface; these currents 
ignite internal supercurrents distribution. Substituting
the expression for the current \req{th-xyz} via fluxes \req{warb}, 
and subsequently,
use the relation \req{phi-ijk} between the fluxes and phases, 
we arrive at the closed system of $3N(N+1)^2$ equations 
for $\tilde{\theta}^{\alpha}_{i,j,k}(t)$ of a homogeneous Josephson array
whose solution will be discussed 
first numerically, and then in physical terms. 
For that purpose we have to introduce first the 
random variations of the junction parameters 
to make the situation more realistic.

\subsection{Inclusion of the grain size distribution and 
            the numerical procedure}

The system of nonlinear equations \req{phi-ijk}, \req{th-xyz} and \req{warb}
is solved numerically and, subsequently, analyzed in
qualitative terms. The physical situation
is reflected by assuming that we have a microwave 
field of period $T$ and amplitude $A_m$ 
applied along the $x$-axis and $x$-axis and a static magnetic field
applied in the $z$-direction. In other words, the 
external magnetic flux piercing the network has the components
  \begin{equation}
  \label{fiext}
    \left\{\begin{array}{c}
      \tilde{\Phi}^{ext}_{x}(\tilde{t})=
        A_m\sin\left(\duzyfrac{2\pi\tilde{t}}{T}\right), \\
      \tilde{\Phi}^{ext}_{y}(\tilde{t})=0, \\
      \tilde{\Phi}^{ext}_{z}(\tilde{t})=\alpha\tilde{t}, 
    \end{array}\right. 
  \end{equation}
where $\alpha<<1$ describes the slow 
sweeping rate of the applied (quasistatic) field. 

The description of real systems requires 
taking into account the random variation of the 
parameters of each loop in the array.
For this purpose we introduce coefficients $a^{xy}_{i,j,k}$, 
$a^{yz}_{i,j,k}$, and $a^{zx}_{i,j,k}$, which characterize the size 
of elementary loop with coordinates (i,j,k). 
These parameters scale the magnetic flux 
$\tilde{\Phi}^{ext}_{x}$ in the following way
  $$
    \tilde{\Phi}^{ext}_{x}(\tilde{t})\rightarrow
    \frac{\tilde{\Phi}^{ext}_{x}(\tilde{t})}{a^{yz}_{i,j,k}}, 
  $$
and in a similar manner the other components.
Analogously, the junction parameters 
$\tilde{L}$ and $\tilde{\Gamma}$ fluctuate and depend on the direction,
e.g. 
  $$
    \tilde{\Gamma}\rightarrow\tilde{\Gamma}^{x}_{i,j,k},\ \ \ \ \ \ 
    \tilde{L}\rightarrow\tilde{L}^{yz}_{i,j,k},\ \ \ \ 
    \mbox{etc.}
  $$
All the parameters fluctuate according 
to the Gaussian distribution 
with $10\%$ dispersion to match the experimentally 
estimated dispersion of crystallite size \cite{stan} arround 
the values $a=a_0\approx 1$, $\tilde{L}_0$, and $\tilde{\Gamma}_0$. 
The representative values of the parameters in both physical and 
dimensionless units (used in the simulation) are provided 
in Table~\ref{tab1}. The microwave field period $\tilde{T}=50$ 
corresponds to the frequency much higher than 
that in the experiments \cite{stan,dur,blaz} ($9.4\mbox{ GHz}$).The
higher frequency was taken to accelerate the 
computations carried out on the work station
Alpha-600~MHz (DEC). We believe that this 
factor does not influence the output character
in any essential way. Besides, the values 
of parameters listed in Table~\ref{tab1} are rough estimates anyway. 

\begin{table}
\caption{
  The estimated values of the sample parameters and the corresponding values 
  of the parameters taken in the network simulation}
\label{tab1}
\begin{tabular}{l|l}
  \hline\hline
  Simulation parameters (dimensionless) & Sample parameters (physical units)\\
  \hline
  Network lattice parameter:         & Average grain size: \\ 
              \hspace{0.5cm} $a_0=1$ &\hspace{0.5cm}$d=0.74\ \mu\mbox{m}$\\
  Microwave field amplitude:         &                          \\
  \hspace{0.5cm} $A_m=0.1$           & 
                      \hspace{0.5cm} $3.6\cdot 10^{-4}\mbox{ T}$ \\
  Applied field range:               &                          \\
  \hspace{0.5cm}$\tilde{\Phi}^{ext}_z=0\div 10$  & 
                       \hspace{0.5cm} $B_z=0\div 0.036\mbox{ T}$ \\  
  Time step:                         &                          \\
  \hspace{0.5cm} $\Delta\tilde{t}=0.1$ & 
                     \hspace{0.5cm} $1.06\cdot 10^{-13}\mbox{s}$ \\
  Microwave field period:            & Microwave frequency  \\
  \hspace{0.5cm} $\tilde{T}=50$      &\hspace{0.5cm} $\nu=190\mbox{ GHz}$ \\
  Relative self-inductance:          & Critical current:      \\
  \hspace{0.5cm}$\tilde{L}\equiv\duzyfrac{LI_c}{\Phi_0}=1.0$ &
      \hspace{0.5cm}
      $j_c=\duzyfrac{I_c}{d^2}=1.7\cdot 10^{10}\mbox{ A/}\mbox{m}^2$ \\
  Damping:                           & Normal state
                                       resistivity: \\ 
  \hspace{0.5cm}$\tilde{\Gamma}=5.0$ & 
    \hspace{0.5cm}
    $\rho_n=R\duzyfrac{d^2}{\Delta}=2.4\cdot 10^{-4}\ \Omega\mbox{m}$\\
   & \\
  \hline\hline
\end{tabular}
\end{table}

The system of $3N(N+1)^2$ differential 
equations, generated for $N=1-30$ was
solved numerically using the fourth-order
Runge-Kutta method \cite{bur} taking randomly 
initial values of $\tilde{\theta}(0)$ and $\dot{\tilde{\theta}}(0)$, 
as well as setting the time step $\Delta\tilde{t}=0.1$. 
The sweeping rate of the quasistatic field was $\alpha=10^{-3}$ 
and in the range $[0,10]$. 
Typically, the solution simulation consisted of $10^5$ time steps.

\section{Analysis of the results}
\subsection{Numerical solution}

\begin{figure*}
\centering\epsfxsize=0.6\textwidth\epsfbox{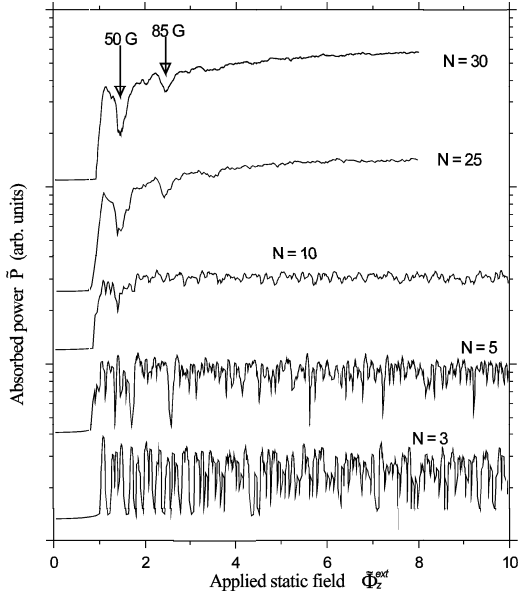}
\caption{
  Power absorption vs applied magnetic field for the network containing
  $N\times N\times N$ junctions; the parameters 
  $\tilde{\Gamma}$ and $\tilde{L}$ where 
  undergoing $10\%$ Gaussian fluctuations around the mean values 
  $\tilde{\Gamma}_0=5$ and $\tilde{L}_0=1$.}
\label{rand3-30}
\end{figure*}

The power absorbed (per one junction) 
can be written in dimensionless units as
  \begin{equation}
  \label{pabs}
    \tilde{P}=
    \frac{1}{3N(N+1)^2}\left<\sum_{i,j,k}
    \tilde{\Gamma}^{x}_{i,j,k}\left(\dot{\tilde{\theta}}^{x}_{i,j,k}\right)^2+
    \tilde{\Gamma}^{y}_{i,j,k}\left(\dot{\tilde{\theta}}^{y}_{i,j,k}\right)^2+
    \tilde{\Gamma}^{z}_{i,j,k}\left(\dot{\tilde{\theta}}^{z}_{i,j,k}\right)^2
    \right>,
  \end{equation}
where averaging takes place over the time interval 
equal to the microwave field period $\tilde{T}$. This expression 
corresponds to the expression $P=U^2/R$ in
dimensionless units, where $U$ is the voltage drop
on resistance $R$.  The representative shape of the
applied field dependence (in units $\Phi/\Phi_0$) 
of $\tilde{P}$ is displayed in \refig{rand3-30} for several
values of $N=3-30$. We see that the discrete
absorption peaks average out with increasing $N$. 
Nonetheless, two characteristic feature survive. 
First of them is that the absorption starts at the value
$\tilde{\Phi}^{ext}_z=0.9$ (corresponding to $10\%$ reduction of
the elementary loop size from the standard value equal 
to unity). Second, well defined minima develop 
at specific values of $\tilde{\Phi}^{ext}_z$. 
These minima are also present on the experimental curve
(cf. \refig{abs-stan}, where the derivative $dP/d\Phi_z$ was measured). 
The first minimum is positioned at 
$\tilde{\Phi}^{ext}_z=1.4$, which in combination with the averange
grain size $d=0.74\ \mu\mbox{m}$ gives the magnetic induction
$B_1=50\mbox{ Gs}$ in an excellent agreement with the experiment. 
The second minimum is placed at $\tilde{\Phi}^{ext}_z=2.4$, 
which leads to the induction value $B_2=85\mbox{ Gs}$.

\begin{figure*}[t]
\centering\epsfxsize=0.6\textwidth\epsfbox{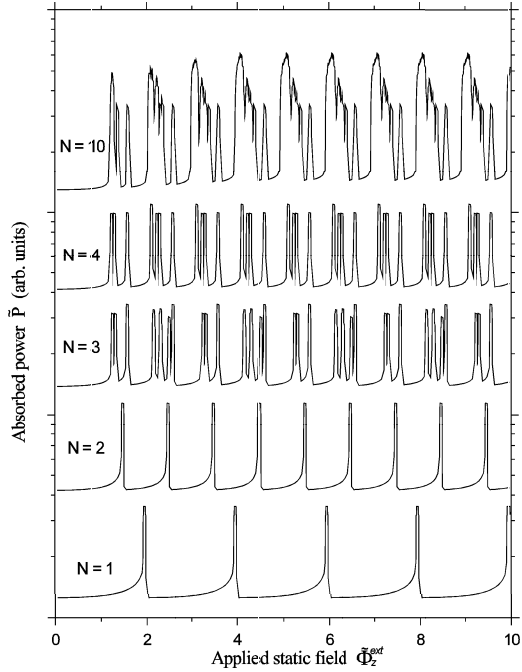}
\caption{
  Applied field dependence of the microwave power absorption 
  for different size of the network of identical Josephson junctions. 
  The parameters are: $\tilde{\Gamma}=5$ and $\tilde{L}=1$.}
\label{reg1-10}
\end{figure*}

The described effect is more pronounced with
increasing $N$ and must be common to all 
inhomogeneous superconductors when the
statistic improves. It is related to the dispersion in size distribution
of the crystallites and should take place only if the statistical
distribution is peaked around one or two sizes (in the second
case it means that we have bicrystallites coexisting with the
crystallites). To understand this secondary features in 
detail have to analyze the array with 
variable $N$ and identical physical parameters. 
The results of the simulation for such network containing
$N\times N\times N$ junctions are diplayed in \refig{reg1-10}, 
for the values of parameters $N = 1 \div 10$, $\tilde{\Gamma}=5$
and $\tilde{L}=1$. 
With increasing $N$ we observe the systematic
filling of the space between the discrete
resonance (physical discussion) of which is provided
below), which eventually for $N\rightarrow\infty$ smoothens out
the absorption curve above the threshold 
value of $\tilde{\Phi}^{ext}_{z}=1$. The situation with a lack
of sizable statistical variations of the junction parameters
may explain the recently observed \cite{niew} smooth variation 
of the power absorption in YBaCuO samples. 

\subsection{Physical discussion}

\begin{figure*}[t]
\centering\epsfxsize=0.6\textwidth\epsfbox{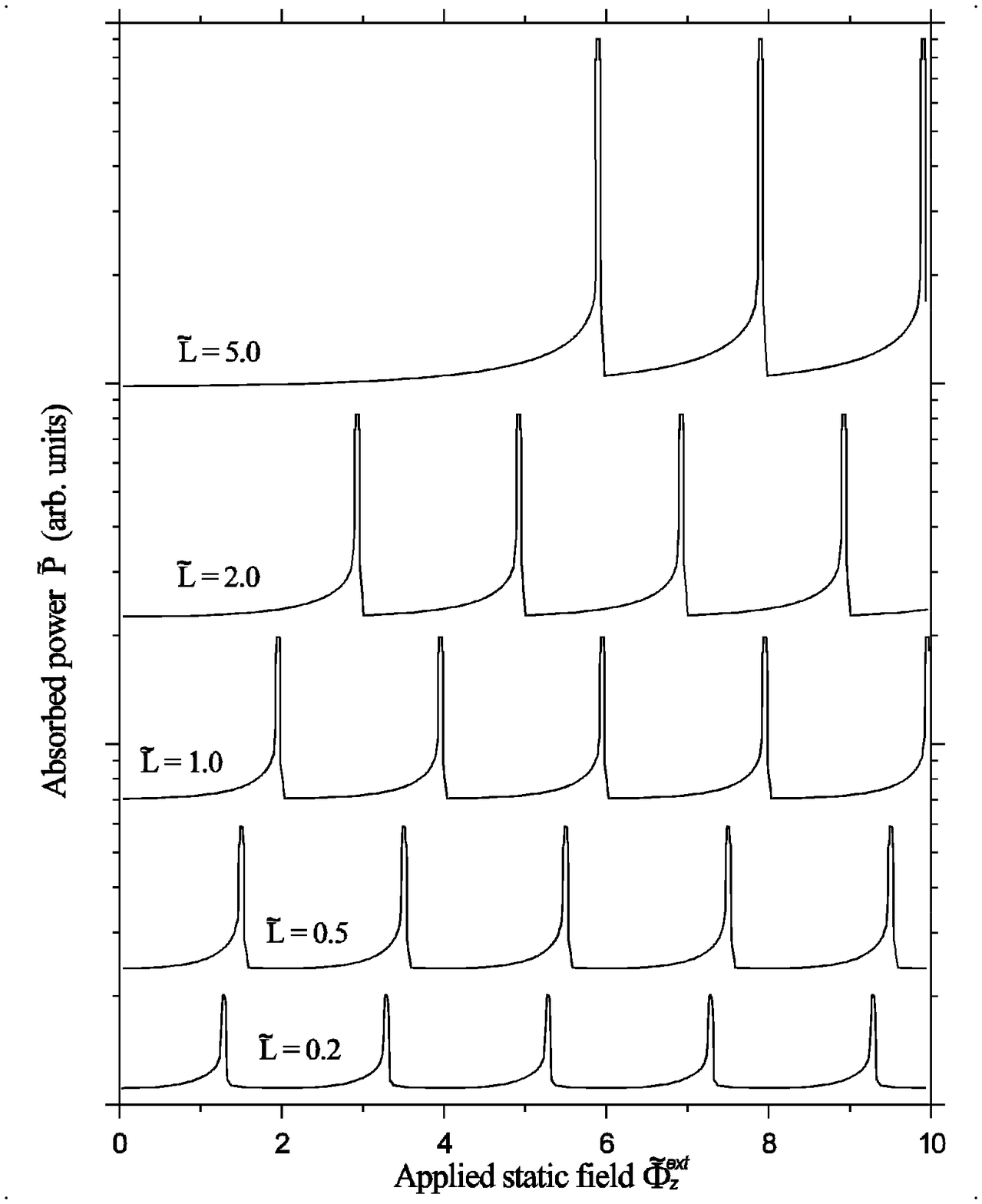}
\caption{
  Microwave absorption in a cube containing 12
  junctions (a structural unit) for the values
  of $\Gamma=5$ and different $\tilde{L}$.}
\label{jj12}
\end{figure*}

The first simulation cycle of the absorption was carried
out on a single cube containing 12 
junctions without random variation of the
parameters. Strictly speaking, we have studied the 
evolution of the absorption maxima with increasing
inductance $\tilde{L}$. Elementary analysis suggests that 
the first absorption maximum appears for a single junction at
$\tilde{\Phi}^{ext}_z=\tilde{L}$, which determines the connection 
between $\tilde{L}$ and the lower critical field for the sample
  \begin{equation}
  \label{Bc1}
    B_{c1}=\frac{\tilde{L}\Phi_0}{d^2}. 
  \end{equation}
This result follows from the circumstance that
for $\tilde{\Phi}^{ext}_z=\tilde{L}$ the current screening the external
flux in $xy$ plane achieves a critical value for
the appearance of nonzero voltage. 
The results presented in \refig{jj12} prove that the
maxima in question appear at higher field
depending. The presence of the microwave
field does not account for the difference,
as it shifts the maximum the opposite way. 

\begin{figure*}[t]
\centering\rotatebox{270}{\epsfysize=0.8\textwidth\epsfbox{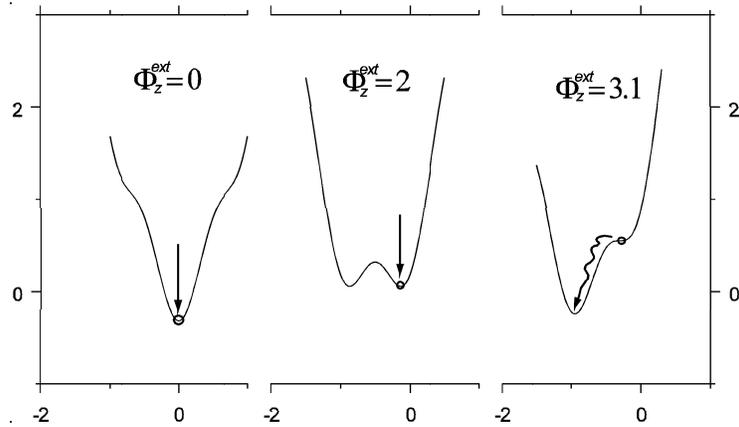}}
\caption{
  Evolution of the effective potential for square loop composed of 4 Josephson
  junctions. The arrow points to the equlibrum state.}
\label{katai}
\end{figure*}

\begin{figure*}[t]
\centering\epsfxsize=0.6\textwidth\epsfbox{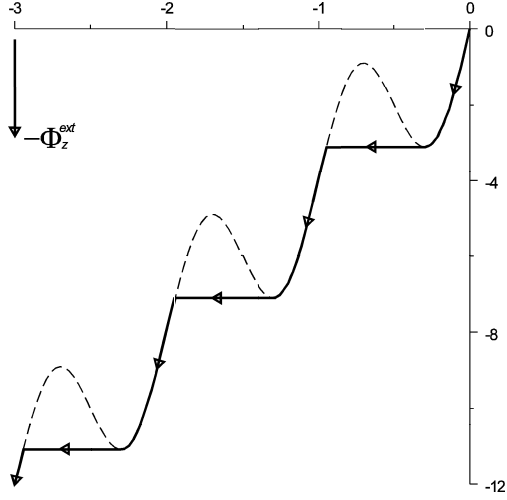}
\caption{
  Solution of Eqn. \req{j4min} for $\tilde{\theta}=0$ and 
  $\tilde{\Phi}^{ext}_z=0$.}
\label{kataii}
\end{figure*}

The qualitative analysis of the differential
equations describing the system helps to explain
the maxima positions in the following manner. 
Let us consider a planar (xy) loop
containing $4$ junctions in the static field $\tilde{\Phi}^{ext}_z$. 
Because of symmetry the Eqs. \req{th-xyz} and \req{phi-ijk} (with
the condition \req{warb}) reduce to the form 
  \begin{equation}
  \label{j4e1}
    \ddot{\tilde{\theta}}+\tilde{\Gamma}\dot{\tilde{\theta}}+
    \tilde{L}\sin 2\pi\tilde{\theta}=
    -4\tilde{\theta}-\tilde{\Phi}^{ext}_{z}, 
  \end{equation}
or equivalently
  \begin{equation}
  \label{j4e2}
    \ddot{\tilde{\theta}}+\tilde{\Gamma}\dot{\tilde{\theta}}=
    -\frac{d}{d\tilde{\theta}}
    \left(
      2\left(
        \tilde{\theta}+\frac{1}{4}\tilde{\Phi}^{ext}_{z}
      \right)^2
      -\frac{\tilde{L}}{2\pi}\cos 2\pi\tilde{\theta}
    \right). 
  \end{equation}
The last equation describes a damped oscillations 
in a potential field
  \begin{equation}
  \label{j4pot}
    V(\tilde{\theta})=
    2\left(
      \tilde{\theta}+\frac{1}{4}\tilde{\Phi}^{ext}_{z}
    \right)^2  
    -\frac{\tilde{L}}{2\pi}\cos 2\pi\tilde{\theta}.   
  \end{equation}
For the absence of applied field the system 
has a minimum at rest: $\tilde{\theta}=0$. With increasing
$\tilde{\Phi}^{ext}_z$ the equilibrium position shifts until
the minimum becomes an inflection point, at which
a catastrophe of $A_2$ type occurs. The situation is
represented schematically in \refig{katai}. The
minima at the $V(\tilde{\theta})$ curves are determined
by the condition
  \begin{equation}
  \label{j4min}
    4\tilde{\theta}+
    \tilde{L}\sin 2\pi\tilde{\theta}=-\tilde{\Phi}^{ext}_{z}.
  \end{equation}
For large $\tilde{L}$ this equation has many solutions, but
starting from $\tilde{\theta}=0$ for $\tilde{\Phi}^{ext}_z$ we can
establish unambiguously, which is realized, as
illustrated in \refig{kataii}. From this Figure we see
that the catastrophes take place for the $\tilde{\theta}$ values, 
for which the function
  \begin{equation}
  \label{j4f}
    f(\tilde{\theta})=4\tilde{\theta}+\tilde{L}\sin 2\pi\tilde{\theta} 
  \end{equation}
acquires minima, i.e. for
  \begin{equation}
  \label{j4thmin}
    \tilde{\theta}=\tilde{\theta}_c=
    -\frac{1}{4}-\frac{1}{2\pi}
    \arcsin\frac{2}{\pi\tilde{L}}+k, 
  \end{equation}
where $k=0,1,2,..$~. The values $\tilde{\theta}_c$ lead to the
values of the flux
  \begin{equation}
  \label{j4fic}
    \tilde{\Phi}^{ext}_z=
    \tilde{\Phi}_c(A_m=0)=-f(\tilde{\theta}_c)=1+\frac{2}{\pi}
    \arcsin\frac{2}{\pi\tilde{L}}+\tilde{L}
    \sqrt{1-\frac{4}{\pi^2\tilde{L}^2}}+4k. 
  \end{equation}
This expression determines approximately only every
second absorption maximum position displayed
in \refig{jj12}. This is because we have ignored the
effect of the microwave field, which is
parallel to the $x$-axis, which breaks the equivalence
between $x$ and $y$ axes. Additional simulation
has shown that even for a very small amplitude
$A_m=10^{-15}$ of the microwave field the picture
is changed drastically: the number of maxima
doubles and their positions shifts by a small
jump. Additional problem is connected with 
the fact that Eq. \req{j4fic} is not defined for $\tilde{L}<2/\pi$
a circumstance leading to a continuos evolution. 
A detailed analysis of stability of the system 
of differential equations describing the discussed
$12$ junctions is quite lengthy and after a linearization 
with respect to $A_m$ leads to the following 
expression replacing \req{j4fic} (see {\em Appendix~A}): 
  \begin{equation}
  \label{j12fic}
    \tilde{\Phi}(A_m\rightarrow 0)=
    1+\frac{2}{\pi}\arcsin\frac{1}{\pi\tilde{L}+1}+
    \tilde{L}\sqrt{1-\frac{1}{(\pi\tilde{L}+1)^2}}+2k. 
  \end{equation}
This approximate analytic result is compared 
with the simulations in Table~\ref{tab2}. 
The agreement is quite good, particularly for large $\tilde{L}$.

\begin{table}
\caption{
  Self-inductance dependence of the first absorption maximum position
  for a cube composed of 12 Josephson junctions 
  (obtained from Eqn. \req{j12fic}),
  and compared with the computed values for a~very small ($10^{-15}$)
  and realistic ($0.1$) amplitudes of the microwave field. }
\label{tab2}
\begin{tabular}{l|l|l|l}
  \hline\hline
  $\tilde{L}$ & Eqn. \req{j12fic} & $A_m=10^{-15}$ & $A_m=0.1$\\
  \hline
  0.2 & 1.579 & 1.616 & 1.276 \\
  0.5 & 1.715 & 1.733 & 1.489 \\
  1.0 & 2.126 & 2.153 & 1.935 \\
  2.0 & 3.069 & 3.093 & 2.906 \\
  5.0 & 6.029 & 6.032 & 5.882 \\
  \hline\hline
\end{tabular}
\end{table}

Stability analysis complicates
enormously for large $N$, even for $A_m=0$. 
However, we can grasp the difference between
the results for $N=1$ and $N=2$ in \refig{reg1-10}. 
Namely, taking for $N=2$ the problem
symmetry then we can reduce effectively our reasoning
to the $N=1$ situation by merely rescaling
the quantities: 
$\tilde{\theta}\rightarrow 2\tilde{\theta}$,
$\tilde{\Phi}\rightarrow 4\tilde{\Phi}$, and 
$\tilde{L}\rightarrow 4\tilde{L}$. 
This explains the doubling of the number of
maxima  when moving from $N=1$ to $N=2$ case. 
The existence of additional maxima for $N>2$
in \refig{reg1-10} indicates the presence of other
oscillation modes leading to the subsidiary
maxima near the main maxima. For
$N=10$ we observe already an absorption
band widening with growing system size. 
Additionally, for $N>2$ a gradual growth of maxima
in the lowest regime of $\tilde{\Phi}^{ext}_z$ takes place. 
The absorption maxima become periodic with the interval of 
one quantum (i.e. $\tilde{\Phi}=1$) only for
$\tilde{\Phi}^{ext}_z>4$ for $\tilde{L}=1$. This phenomenon 
seems to be analogous to the existence of two critical fields
in bulk samples. 

The character change of the
absorption spectrum obtained after introducing
the random fluctuations of the junction parameters
is also easy to grasp quantitatively. Each discrete absorption
line for a regular array depicted in \refig{reg1-10} is convoluted
with the probability distribution. The lines for large
$\tilde{\Phi}^{ext}_z$ are smeared out proportionally in a wide region, 
so their sum will result in a practically constant value of the absorption.
The expression \req{Bc1} for the first critical
field is valid; that is why it was used to the
evaluation of the quantities listed in Table~\ref{tab1}. 
This is simply due to the fact 
that for $\tilde{\Phi}^{ext}_z<\tilde{L}$ absorption is zero, since
the superconducting loops screen the external field. 

As we have seen, the absorption
effects are amplified with growing $N$. This proves
that they are common to all inhomogeneous superconductors, 
in distinction to the dispersion of results
in one simulation cycle, which disappears with the 
better statistics.

\section{Summary}

A general method has been presented of generating a
system of second-order differential equations, which
describe the time evolution of three-dimensional
Josephson networks with inclusion of their
resistivity, inductance and capacity. The effect of junction
capacity is important, since we include high frequency
microwave field. Apart from a characteristic field
$B_{c1}$ we observe the existence of second critical 
field described above. 

The analysis of the stability of those equations
for a single structural cell (a cube containing
$12$ junctions) provides the absorption maxima
positions only after inclusion of the small microwave
field amplitude. Also, the
networks of the size $30\times 30\times 30$ and with a
$10\%$ random variations of the junction parameters
are sufficient to account for the 
experimental observations.

\section*{Acknowledgement}

The authors acknowledge the support
of the KBN grant No. 2P03B~129~12. 
A.~R. acknowledges a special grant of Ministry of
Education of Poland for undergraduate students. 
The discussion with our colleagues: Prof. A.~Ko{\l}odziejczyk,
J.~Niewolski and Prof. J.~Stankowski are appreciated.

\appendix
\renewcommand{\theequation}{\Alph{section}\arabic{equation}}
\setcounter{equation}{0}
\section{Stability of the system of differential
          equations for the cubic elementary cell of junctions}

\begin{figure*}
\centering\epsfxsize=0.6\textwidth\epsfbox{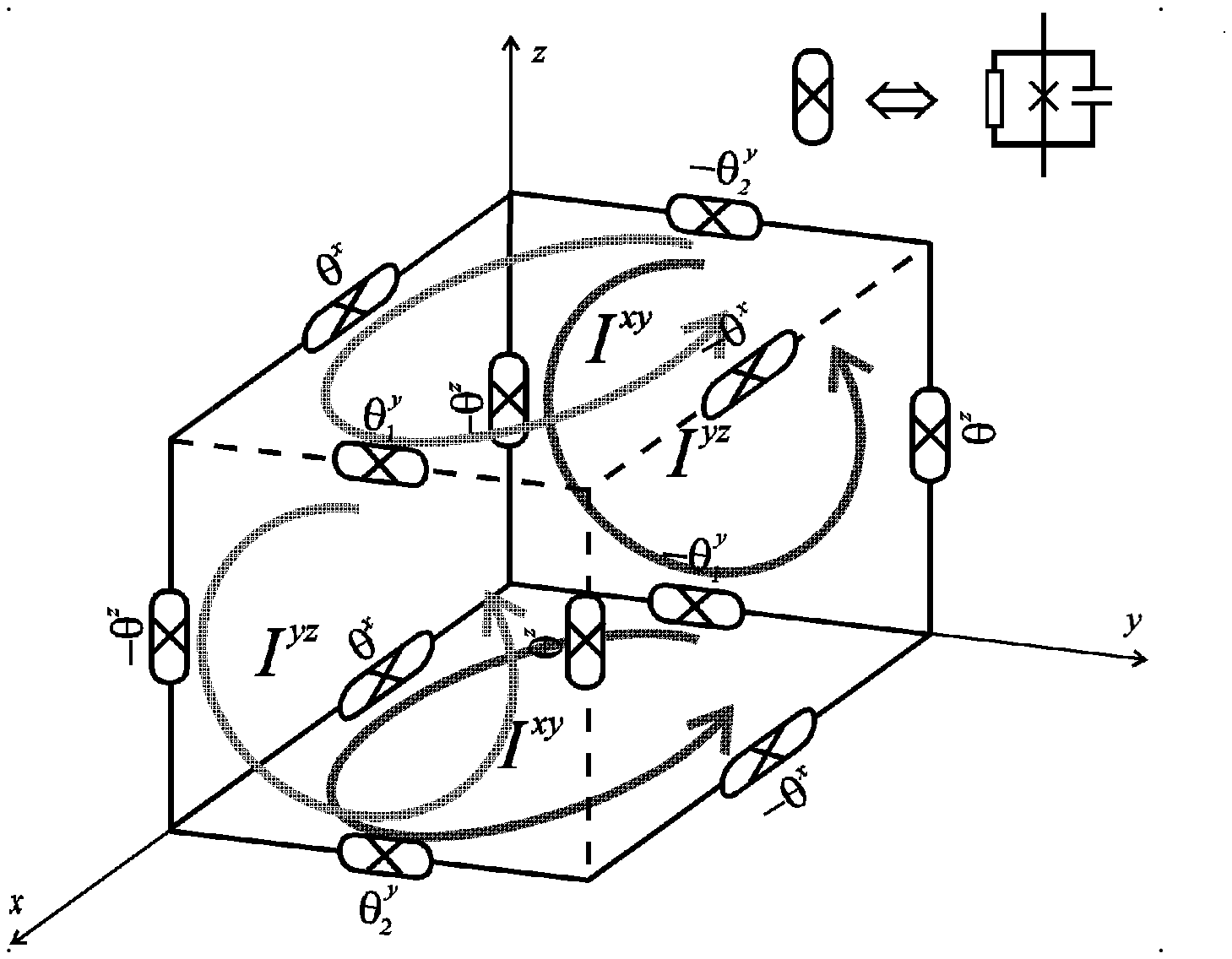}
\caption{
  Isolated cubic elementary cell composed of $12$ Josephson junctions,
  together with the corresponding independent gauge invariant 
  phases and currents. The independent magnetic fluxes 
  reflect the marked currents.}
\label{kostka}
\end{figure*}

In this Appendix we derive the analytic result 
\req{j12fic} from the stability considerations for the system
of differential equations, describing the time evolution
of the phases. The reasoning is carried out for an 
isolated structural unit composed of $12$ 
Josephson junctions (an elementary cube) with inclusion of 
resistances and capacitances. 
This cell is placed in the applied field static parallel 
to $z$-axis and the microwave field parallel to $x$-axis. 
Then, the symmetry of the problem reduces to consideration of 
four invariant phases $\theta^x$, $\theta^z$, $\theta^y_1$, 
and $\theta^y_2$, as well as of two magnetic fluxes $\Phi^{xy}$
and $\Phi^{yz}$, and to the loop current $I^{xy}$ and $I^yz$. The
situation is depicted schematically in \refig{kostka}.
The magnetic fluxes can be related to the 
phase differences and the currents in the following 
manner (cf. Eqs. \req{bezwym2})
  \begin{equation}
  \label{kostka-fi}  
    \left\{\begin{array}{c}
      \tilde{\Phi}^{xy}= - \left(\tilde{\theta}^y_2 + \tilde{\theta}^y_1 
                         + 2\tilde{\theta}^x \right) 
                       = \tilde{\Phi}^{ext}_z + \tilde{L}\tilde{I}^{xy}, \\
      \tilde{\Phi}^{yz}= - \left(\tilde{\theta}^y_2 - \tilde{\theta}^y_1 
                         + 2\tilde{\theta}^z \right)  
                       = \tilde{\Phi}^{ext}_x + \tilde{L}\tilde{I}^{yz}, \\
    \end{array}\right. 
  \end{equation}
where $\tilde{\Phi}^{ext}_x$ and $\tilde{\Phi}^{ext}_z$ are the fluxes
parallel to the corresponding axes (in units of $\Phi_0$). 

The dynamic equations \req{th-xyz} take the form: 
  \begin{equation}
  \label{kostka-th1}
    \left\{\begin{array}{c}
      \ddot{\tilde{\theta}}^x + \tilde{\Gamma}\dot{\tilde{\theta}}^x 
        + \tilde{L}\sin 2\pi\tilde{\theta}^x
      = \tilde{L}\tilde{I}^{yz} ,\\
      \ddot{\tilde{\theta}}^y_1 + \tilde{\Gamma}\dot{\tilde{\theta}}^y_1 
        + \tilde{L}\sin 2\pi\tilde{\theta}^y_1 
      = \tilde{L}\left(\tilde{I}^{xy}-\tilde{I}^{yz}\right) ,\\
      \ddot{\tilde{\theta}}^y_2 + \tilde{\Gamma}\dot{\tilde{\theta}}^y_2 
        + \tilde{L}\sin 2\pi\tilde{\theta}^y_2 
      = \tilde{L}\left(\tilde{I}^{xy}+\tilde{I}^{yz}\right) ,\\
      \ddot{\tilde{\theta}}^z + \tilde{\Gamma}\dot{\tilde{\theta}}^z 
        + \tilde{L}\sin 2\pi\tilde{\theta}^z 
      = \tilde{L}\tilde{I}^{xy} .\\
    \end{array}\right. 
  \end{equation}

Adding and substracting second and third 
equations in \req{kostka-th1}, and substituting expressions 
for $\tilde{L}\tilde{I}^{xy}$ and $\tilde{L}\tilde{I}^{yz}$ 
calculated from \req{kostka-fi}, we obtain
  \begin{equation}
  \label{kostka-th2}
    \left\{\begin{array}{c}
      \ddot{\tilde{\theta}}^x + \tilde{\Gamma}\dot{\tilde{\theta}}^x 
        + \tilde{L}\sin 2\pi\tilde{\theta}^x 
      = -2\tilde{\vartheta}_+ -2\tilde{\theta}^x -\tilde{\Phi}^{ext}_z, \\
      \ddot{\tilde{\vartheta}}_+ + \tilde{\Gamma}\dot{\tilde{\vartheta}}_+
        + \tilde{L}\cos 2\pi\tilde{\vartheta_-}\sin 2\pi\tilde{\vartheta}_+ 
      = -2\tilde{\vartheta}_+ -2\tilde{\theta}^x -\tilde{\Phi}^{ext}_z, \\
      \ddot{\tilde{\vartheta}}_- + \tilde{\Gamma}\dot{\tilde{\vartheta}}_-
        + \tilde{L}\cos 2\pi\tilde{\vartheta_+}\sin 2\pi\tilde{\vartheta}_- 
      = -2\tilde{\vartheta}_- -2\tilde{\theta}^z -\tilde{\Phi}^{ext}_x, \\
      \ddot{\tilde{\theta}}^z + \tilde{\Gamma}\dot{\tilde{\theta}}^z 
        + \tilde{L}\sin 2\pi\tilde{\theta}^z 
      = -2\tilde{\vartheta}_- -2\tilde{\theta}^z -\tilde{\Phi}^{ext}_x, \\
    \end{array}\right. 
  \end{equation}
where the new variables are defined as
  \begin{equation}
  \label{def-varth}
    \tilde{\vartheta}_+\equiv
      \frac{1}{2}(\tilde{\theta}^y_2+\tilde{\theta}^y_1), 
    \ \ \ \ \ \ 
    \tilde{\vartheta}_-\equiv
      \frac{1}{2}(\tilde{\theta}^y_2-\tilde{\theta}^y_1). 
  \end{equation}

The last two equations in \req{kostka-th2} describe 
damped oscillations of two coupled nonlinear oscillators
under the influence of the external force
  \begin{equation}
  \label{kostka-F}
    \underline{F}=\left(\begin{array}{c} F_{\tilde{\vartheta}_-} \\ 
                                   F_{\tilde{\theta}^z}  \\
            \end{array}\right), 
    \ \ \ \ \ \ 
    F_{\tilde{\vartheta}_-}= F_{\tilde{\theta}^z}= -\tilde{\Phi}^{ext}_{x}; 
  \end{equation}
and in the potential 
  \begin{equation}
  \label{kostka-V}
    V\left(\tilde{\vartheta}_-, \tilde{\theta}^z\right)=
    -\frac{\tilde{L\cos 2\pi \tilde{\vartheta}_+}}{2\pi}
    \cos 2\pi\tilde{\vartheta}_- +\left(\tilde{\vartheta}_-\right)^2 
    +2\tilde{\vartheta}_-\tilde{\theta}^z +\left(\tilde{\theta}^z\right)^2
    -\frac{\tilde{L}}{2\pi}\cos 2\pi\tilde{\theta}^z.  
  \end{equation}

We show below that the $\tilde{\vartheta}_+$ dependence
of the potential is connected to $\tilde{\Phi}^{ext}_z$ and 
thus vanishes if $\tilde{\Phi}^{ext}_z=0$. Therefore, 
$V\left(\tilde{\vartheta}_-, \tilde{\theta}^z\right)$ 
has a sharp minimum 
at $\tilde{\vartheta}_-=\tilde{\theta}^z=0$. Furthermore, since along
the $x$-axis $\tilde{\Phi}^{ext}_x=A_m\sin\left(2\pi\tilde{t}/T\right)$,
with $A_m<<1$, and the force is given by \req{kostka-F}, we have for small
vibrations around minimum that
  \begin{equation}
  \label{kostka-Vlin}
    V\left(\tilde{\vartheta}_-, \tilde{\theta}^z\right)\cong
    \frac{1}{2}\left(\begin{array}{cc}
                     \tilde{\vartheta}_- & \tilde{\theta}^z \\
               \end{array}\right)
    \left(\begin{array}{cc}
      2\pi\tilde{L}\cos 2\pi\tilde{\vartheta}_+ +2 & 2 \\
      2 & 2\pi\tilde{L}+2 \\
    \end{array}\right)
    \left(\begin{array}{c}
      \tilde{\vartheta}_- \\ \tilde{\theta}^z \\
    \end{array}\right) .
  \end{equation}
The instability threshold of the system \req{kostka-th2}
is reached when the quadratic form \req{kostka-Vlin} 
ceases to be positively determined. This 
happens for
  \begin{equation}
  \label{kryt-varth}
    \tilde{\vartheta}_+=-\frac{1}{4}
    -\frac{1}{2\pi}\arcsin\frac{1}{\pi\tilde{L}+1}. 
  \end{equation}
It remains still to determine the relation
between $\tilde{\vartheta}_+$ and $\tilde{\Phi}^{ext}_z$. 
Since $A_m<<1$, then we can set at $t=0$ $\tilde{\Phi}^{ext}_z=0$; then
  $$
    \tilde{\theta}^y_1\cong\tilde{\theta}_y^2\cong\tilde{\theta}^x, 
  $$ 
and hence
  \begin{equation}
  \label{przybl-varth}
    \tilde{\vartheta}_+\cong\tilde{\theta}^x, \ \ \ \ \ \ 
    \tilde{\vartheta}_-\cong 0. 
  \end{equation}
Thus, the relation to $\tilde{\Phi}^{ext}_z$ can be determined 
by considering the equilibrium points, for which 
e.g. the first equation of \req{kostka-th2} reduces to
the form
  $$
    \tilde{L}\sin 2\pi\tilde{\theta}^x +2\tilde{\vartheta}_+ 
    +\tilde{\theta}^x= -\tilde{\Phi}^{ext}_z. 
  $$
Taking into account \req{przybl-varth} and \req{kryt-varth} 
we obtain the critical flux value
  \begin{equation}
  \label{kryt-Phi}
    \tilde{\Phi}_c= 1+\frac{2}{\pi}\arcsin\frac{1}{\pi\tilde{L}+1}
    +\tilde{L}\sqrt{1-\frac{1}{(\pi\tilde{L}+1)^2}}.
  \end{equation}
This formula determines the applied field 
value, for which the first catastrophe takes place. 
Formally, the critical value of $\tilde{\vartheta}_+$, for which
the determinant of the quadratic form \req{kostka-Vlin} 
vanishes, repeats every unit. In effect, the
critical value $\Phi_c$ should repeat itself every
$4$ units. This does not happen because at 
the catastrophe, at which \req{kostka-V} changes locally 
from paraboloidal into saddle-like, the 
small oscillation approximation looses its meaning.
The $\tilde{\vartheta}_-$ value does not grow without limit
but changes abruptly by the amount $\pm\frac{1}{2}$, 
i.e. the system jumps to the closest
minimum, where again we can use for
the time being the small-oscillation
considerations. The sign is not important. 
Nonetheless, such change of $\tilde{\vartheta}_-$ 
determines the sign change of the term
$\cos\left(2\pi\tilde{\vartheta}_-\right)$ 
in the second equation of \req{kostka-th2} system. 
Therefore,  the equilibrium condition for 
$\tilde{\theta}^x$ and $\tilde{\vartheta}_+$ 
has now the form
  $$
    \left\{\begin{array}{c}
      \tilde{L}\sin 2\pi\tilde{\theta}^x +2\tilde{\vartheta}_+ 
      +2\tilde{\theta}^x= -\tilde{\Phi}^{ext}_z, \\
      -\tilde{L}\sin 2\pi\tilde{\vartheta}_+ +2\tilde{\vartheta}_+ 
      +2\tilde{\theta}^x= -\tilde{\Phi}^{ext}_z, \\
    \end{array}\right. \ 
  $$ 
from which we have
  $$
    \sin 2\pi\tilde{\theta}^x +\sin 2\pi\tilde{\vartheta}_+ = 0,
  $$
and thus, the magnitude of the jump is
  $$
    \tilde{\vartheta}_+\cong\tilde{\theta}^x\rightarrow 
    \tilde{\vartheta}_+\cong\tilde{\theta}^x\pm\frac{1}{2}.  
  $$
This means that at each catastrophe the
system overcomes, at given magnetic field, 
half of the way to the next critical point 
(or, pulls back by the same value). This, 
in turn leads to the frequency doubling, 
with which catastrophes happen with a slow
increase of static external field. In effect
this leads to
  \begin{equation}  
  \label{kryt-Phi2}
    \tilde{\Phi}_c= 1+\frac{2}{\pi}\arcsin\frac{1}{\pi\tilde{L}+1}
    +\tilde{L}\sqrt{1-\frac{1}{(\pi\tilde{L}+1)^2}}+2k,
  \end{equation}
where $k=0,1,2,...\ $. This is formula \req{j12fic} in the main text.


\end{document}